\documentclass[12pt]{article}
\newcommand{\beao}{\begin{eqnarray*}}
\newcommand{\eeao}{\end{eqnarray*}}
\usepackage{a4}\usepackage{amsfonts}\textwidth15cm
\newcommand{\be}{\begin{equation}}
\newcommand{\ee}{\end{equation}}
\newcommand{\bea}{\begin{eqnarray}}
\newcommand{\eea}{\end{eqnarray}}
\newcommand{\beq}{\begin{eqnarray}}
\newcommand{\eeq}{\end{eqnarray}}
\newcommand{\nn}{\nonumber}
\newcommand{\pa}{\partial}
\newcommand{\ep}{\epsilon}

\newcommand{\om}{\omega}
\newcommand{\Ref}[1]{(\ref{#1})}
\newcommand{ \text}[1]{#1 }
\newcommand{\tr}{{\rm tr}}
\begin{document}
\title{The Casimir effect for a sphere and a cylinder in front of  plane and corrections to the proximity force theorem}
\author{
{\sc M. Bordag}\thanks{e-mail: Michael.Bordag@itp.uni-leipzig.de} \\
\small  University of Leipzig, Institute for Theoretical Physics\\
\small  Vor dem Hospitaltore 1, 04103 Leipzig, Germany}
\maketitle
 \begin{abstract}
Using a path integral approach we rederive a recently found representation of the Casimir energy for a sphere and a cylinder in front of a plane and derive the first correction to the proximity force theorem.
  \end{abstract}

 \section{Introduction}

In view of the rapid experimental progress in the Casimir force measurements and increasing interest in the field of nanotechnology there is a growing demand for reliable theoretical predictions. 
Beyond geometries allowing for a separation of variables one was left so far basically with the proximity force approximation for ideal reflecting surfaces or with the pairwise summation for dilute dielectric bodies.  Also the basic formula for the interaction of dielectric bodies, the Lifshitz formula, is known for simple geometries only (for a review see \cite{Bordag:2001qi}).
There is a number of approximate methods trying to overcome these limitations. The oldest one is the semiclassical approach (see, for example the recent papers \cite{Schaden:1998gq,Schaden:2005mu} and references therein) which  exists in several modifications including the optical path approximation \cite{Scardicchio:2004fy}. Another method is the multiple reflection approximation which recently has been reconsidered in the context of the Casimir effect \cite{Balian:2004jv}. New developments are the multi-scattering approach  of \cite{Bulgac:2005ku} which was applied to the geometry of two spheres and of a sphere in front of a plane and the calculation in \cite{Emig:2006uh} for a cylinder and a plane using a path integral formulation together with a trace formula.  Significant progress had been reached in the world line method  \cite{Gies:2006bt} which is now able to provide corrections to the small separation behavior.

In the present paper we use the path integral approach developed in \cite{Bordag:1985zk} and \cite{Robaschik:1986vj} and apply it to the geometry of a sphere  and a cylinder in front of a plane for Dirichlet boundary conditions. We rederive the formulas obtained in \cite{Bulgac:2005ku} and \cite{Emig:2006uh} in a simpler way and calculate the first correction to the proximity force theorem. 

The paper is organized as follows. In the next section we describe the method of \cite{Bordag:1985zk} and \cite{Robaschik:1986vj} in detail. In the third section we apply the method to  the geometry of a sphere in front of a plane. In the fourth section we discuss some other geometries. Conclusions are given in the final section. Throughout the paper we put $\hbar=c=1$

\section{Functional integration and boundary conditions}
A simple, but powerful method to introduce boundary conditions into a quantum field theory 
is to use functional delta functions. This method was introduced in \cite{Bordag:1985zk} for the calculation of the radiative corrections to the Casimir effect for parallel plates in a covariant gauge. We present it here in the simpler case for a scalar field theory with Dirichlet boundary conditions. The basic idea of the method is to restrict the space of fields over which the functional integration runs to such that fulfil the boundary conditions.   Technically this can be achieved by representing the generating functional of the Green's functions as
\be
\label{Z1}Z(J)=\int D\phi \ \prod_{x\in S}\delta(\phi(x)) \ exp\{- S \}
\ee
with an action
\be\label{S1}S= \int dx \ \left( \frac12 \phi(x)\left(-\pa^2+m^2\right)\phi(x)+J(x)\phi(x)\right),
\ee
where we included  the term with the source  $J(x)$. This is the usual action for a scalar field with mass $m$.
  For simplicity we write all formulas in the Euclidean formulation.

The functional $Z(J)$, \Ref{Z1}, taken without the delta function  is the  generating functional of the corresponding field theory without boundary conditions. Obviously, in \Ref{Z1} more complicated actions can be inserted, especially such which include interactions. However, in connection with the Casimir effect we restrict ourselves here to a free field theory, i.e., to a quadratic action. 
Also we mention that the initial space where the field $\phi$ is defined does not need to be flat or free of boundaries. The method to implement boundary conditions through functional delta functions is completely general, at last on the formal level taken in this section.

The next step in the method is the use of a Fourier representation for  the delta function,
\be\label{Fourier}
\prod_{x\in S}\delta(\phi(x))=\int Db \ exp\left(i\int_S dz \ b(z) \phi(f(z))\right),
\ee
where we described the surface $S$ by a function $f(z)$ such that $x=f(z)\in S$ and the coordinates $z$ are a suitable parameterization of $S$. In fact, \Ref{Fourier} is a functional integral over a scalar field $b(z)$ defined on the surface $S$. Using \Ref{Fourier} in the generating functional \Ref{Z1} we represent $Z(J)$ in the form
\be\label{Z2} Z(J)=\int D\phi \ Db \ \exp(-\tilde{S})
\ee
with
\bea\label{S2}\tilde{S}&=&\frac12 \int dx \ \phi(x)\left(-\pa^2+m^2\right)\phi(x)
\nn \\ &&
-i\int dx \ \int_S dz\  \phi(x) H(x,z) b(z)+\int dx  J(x)\phi(x)\ .
\eea
Here we introduced the new notation,
\be\label{H}H(x,z)=\delta(x-f(z))
\ee
projecting the argument $x$ on the surface $S$. Obviously, $\tilde{S}$ is quadratic in the fields $\phi(x)$ and $b(z)$, hence $\tilde{S}$ can be diagonalized,
\bea\label{S3}\tilde{S}&=&\frac12 \int dx \ \left(\phi(x)-\phi_0(x)\right)\left(-\pa^2+m^2\right)\left(\phi(x)-\phi_0(x)\right)
 \nn\\ &&
+\frac12 \int_S dz \int_S dz' \ \left(b(z)-b_0(z)\right) K(z,z') \left(b(z)-b_0(z)\right)
\nn \\ &&
-\frac12 \int dx \int dy \ J(x) \ ^S\!D(x,y) J(y)  
\eea
with 
\bea\label{phi0}\phi_0(x)&=&\int dx' \ D(x,x')
\left(     i\int_S dz \ H(x',z)\ b(z)-J(x')\right),
\nn \\ 
b_0(z)&=&-i\int_Sdz' \int dx \, dx' \ K^{-1}(z,z')H(x,z')D(x,x')J(x'),
\eea
and
\bea\label{K}K(z,z')&=&\int dx \int dy \ H(x,z) D(x,y) H(y,z')\nn \\ &=&
D(f(x),f(y)).
\eea
The function $D(x,y)$ is the inverse of the kernel of the free action,
\be\label{D1}\left(-\pa^2+m^2\right) D(x,y)=\delta(x-y).
\ee
It is just the usual propagator of the free field theory defined by the action $S$, \Ref{S1}. Instead, the new quantity  
\be\label{SD}^S\!D(x,y)=D(x,y)-\int_S dz \int_S dz' \ D(x,f(z)) K^{-1}(z,z') D(f(z'),y),
\ee
which we introduced in \Ref{S3}, 
is the propagator of the given field theory with boundary conditions. Finally, $K^{-1}(z,z')$ is the inverse of $K(z,z')$ on the surface $S$,
\be\label{K-1}\int_s dz' \ K(z,z') K^{-1}(z',z'')=\delta(z-z'').
\ee
This inversion and further operations like the trace or the determinant with $K(z,z')$ are to be understood as done with an operator $\hat{K}$ whose integral kernel in a space of suitable functions defined on the surface $S$ is just $K(z,z')$. 

With Eq.\Ref{K-1} it is easy to check that $^S\!D(x,y)$, \Ref{SD}, is indeed the propagator which fulfills the boundary conditions. First we remark that it is a propagator. Obviously equation \Ref{D1} is fulfilled for $x\notin S$. Inserted into Eq.\Ref{D1}, the first term delivers the necessary delta function. The second term delivers a delta function too which is, however, non vanishing for $x\in S$ only.  This is not a problem because in the presence of boundary conditions the equation has to be fulfilled outside the boundary only. So it remains to check that the boundary conditions are fulfilled. Put $x$ on the surface $S$, i.e., consider $x\in S$. Then in the second term we can apply \Ref{K-1} and we get zero as required by the Dirichlet condition.

The calculation of the inverse on $S$ of K is the main step in the method. All steps before are merely a formal rewriting. The inversion of $K$ on $S$, however, is equivalent to solve the wave equation with boundary conditions and, in this sense, not much progress is archived. 

After rewriting the exponential in the form given by Eq. \Ref{S3}, it is quadratic in the fields. The integrals are Gaussian and after integration we represent $Z(J)$, \Ref{Z1}, in the form
\bea\label{Z2a}Z(J)&=&C \ \left(\det\left(\pa^2+m^2\right)\right)^{-\frac12} \
\left(\det K(z,z')\right)^{-\frac12} \ \nn \\ && \times 
exp\left\{-\frac12 \int dx \ \int d y\ J(x) \ ^S\!D(x,y) J(y) \right\},
\eea
where $C$ is a constant. In this formula the first determinant comes from the integration over $\phi$. It is the usual one appearing also without boundary condition. Of course, it is independent on the boundary conditions.  The second determinant comes from the integration over $b$. It does depend on the boundary conditions. The source term has the standard form and it generates the usual Feynman rules  with the boundary dependent propagator $^S\!D(x,y)$ instead of the usual one. In this way the field theory with boundary conditions was constructed in \cite{Bordag:1985zk}. 

The last step in the method, which was done in \cite{Robaschik:1986vj}, is to consider the free energy
\be\label{F} F=-\frac{1}{T}\ln Z(0),
\ee
where $T$ is the total time which is the logarithm of the generating functional at zero source term. In \cite{Robaschik:1986vj} this was done for a theory with finite temperature, but of course we can use it at zero temperature too. In this way we get for the boundary dependent part of the Casimir energy 
\be\label{Ec}E_{\rm Casimir}=-\frac{1}{T}\ln \left(\det K(z,z')\right)^{-\frac12}
=\frac{1}{2T} \ {\tr} \ln K.
\ee
The trace in this formula is to be taken in the space of functions on the boundary surface $S$ and $K$ is the operator in this space, whose integral kernel is the function $K(z,z')$, defined in Eq.\Ref{K}. In this way the Casimir energy is represented by an expression on the surface $S$. The contribution of the free space, which contains the highest ultraviolet divergences, had been dropped  in \Ref{Ec}. 

Let us mention that the evaluation of the trace in \Ref{Ec} is a problem of the same difficulty as the inversion of $K$ or the solution of the wave equation with boundary condition, and again, so far we have only a formal rewriting of the original problem.

\section{A sphere and a cylinder in front of a plane}
In this section we apply the general formula \Ref{Ec} for the Casimir energy to the geometry of a sphere and of a cylinder in front of a plane. 
We start with the remark that the derivation in the preceding section is valid also if the initial space is not the full $\Bbb R^4$ but only a half of it with $z\le a$ and if the field $\phi(x)$ satisfies Dirichlet (or other) boundary conditions on the plane $z=a$: $\phi\raisebox{-5pt}{$\mid_{z=a}$}=0$. In that case, for the propagator $D(x,x')$, \Ref{D1}, which has to fulfill Dirichlet boundary conditions at $z=a$ we take
\be\label{DD0} D(x,x')\to D_D(x,x')=D(x-x')-D(\tilde{x}-x')\ ,
\ee
where $\tilde{x}=(x,y,2a-z,t)$ is the  coordinate reflected on the plane $z=a$ and
\be\label{}D(x-x')=\int\frac{d^4k}{(2\pi)^4}\ \frac{e^{-ik(x-x')}}{k^2+m^2}
\ee
is the free space propagator. Here $m$ is the mass of the field $\phi$ but in the following we restrict ourselves to massless fields. 

In the next step we use the stationarity of the problem and turn to the Fourier transform in the time coordinate, $x_4 \to \om$. In this way the Casimir energy \Ref{Ec} can be represented as
\be\label{Eom} E=\frac{1}{2\pi}\int_{0}^\infty d\om\ \tr \ln K_\om
\ee
with (cf. \Ref{K}) 
\be\label{Kom}K_\om(z,z')=D_{D,\om}(f(z),f(z'))
\ee
and
\be \label{Ddom}
D_{D,\om}(x,x')=D_\om(x-x')-D_\om(\tilde{x}-x').
\ee
as well as the free space propagator after Fourier transform in the time direction,
\be\label{Dom3}D_\om(x-x')=\int\frac{d^3k}{(2\pi)^3}\ \frac{e^{-ik(x-x')}}{\om^2+k^2} \ .
\ee
In \Ref{Eom} the trace is now only over the surface $S$ and we used the symmetry of $K_\om$ under $\om\to -\om$. We note that the function  $K_\om$, \Ref{Kom} is even in $\om\to -\om$.   We take this into account in all   integration over $\om$ below.

Now let us consider a sphere of radius $R$ centered in the origin so that the distance from its center to the plane is $a$. In terms of spherical coordinates $(r,\theta,\varphi)$ it is described by constant $r$ and the angles provide a parameterization of $S$, $z=(\theta,\varphi)$. A basis of functions in this space are the spherical harmonics $\mid l,m\rangle=Y_{l,m}(\theta,\phi)$. Now we calculate the trace in \Ref{Eom} in this basis, i.e., we represent the energy by 
\be\label{Eps1}E=\frac{1}{2\pi} \int_0^\infty  d\om  \ \tr _{l,m} \ln K_{lm,l'm'}
\ee
with
\bea\label{}K_{lm,l'm'}&=&\langle l,m\mid D_D(x,x') \mid l,m\rangle
\nn \\ &=& \int d\theta_x \sin \theta_x d\varphi_{x} \int d\theta_{x'} \sin \theta_{x'} d\varphi_{x'} Y_{l,m}(\theta_x ,\varphi_{x}) \ D_D(x,x') \ Y_{l'm'}^*(\theta_{x'} ,\varphi_{x'} )\nn
\eea
and $r_x=r_{x'}=R$. Now the trace is over the indices of the infinite dimensional matrix $\ln K_{lm,l'm'}$. An equivalent representation  is in terms of the determinant of this matrix,
\be\label{} \tr \ln K_{lm,l'm'}=\ln \det(K_{lm,l'm'}).
\ee
Next we use the well known explicit expansion   of the free space propagator,
\be\label{}D_\om(x-x')=\sum_{l,m}Y_{l,m}(\theta_x ,\varphi_{x}) d_l(r,r') Y_{l'm'}^*(\theta_{x'} ,\varphi_{x'} ),
\ee
with
\be\label{dl25}d_l(r,r')=\frac{1}{\sqrt{rr'}}j_l(\om r_<)h^{(1)}_l(\om r_>),
\ee
where $j_l(z)$ and $h^{(1)}_l(z)$ are the spherical Bessel functions  to represent $K$ in the form
\be\label{Klls}K_{l,l'}=\delta_{l,l'} \ d_l(R,R)-\langle lm\mid D_\om(\tilde{x}-x')\mid l'm\rangle \ ,
\ee
where because of the  the azimuthal symmetry we could introduce
\be\label{}K_{lm,l'm'}=\delta_{m,m'}K_{l,l'}.
\ee
The last step is to separate a distance independent contribution by writing
\be\label{}\ln K_{l,l'}=\ln \left(\delta_{l,l'} \ d_l(R,R)\right) + \ln \left(\delta_{l,l'}-\frac{1}{d_l(R,R)} \langle lm\mid D_\om(\tilde{x}-x')\mid l'm\rangle\right).
\ee
The distance $a$ enters this expression only trough $\tilde{x}$ in the second term in the r.h.s. whereas the first is independent of $a$. In fact, it is this term which contains the surface divergences which are still present in the Casimir energy given by Eq. \Ref{Eps1}. But dropping it we come to 
\be\label{Eps2}E=\frac{1}{2\pi} \int_0^\infty d\om  \ \tr _{l,m} \ln\left(\delta_{l,l'}-A_{l,l'} \right),
\ee
where we introduced the notation
\be\label{Alls0}A_{l,l'}=\frac{1}{d_l(R,R)}\langle lm\mid D_\om(\tilde{x}-x')\mid l'm\rangle \ .
\ee
The energy $E$, \Ref{Eps2}, is the ultraviolet finite, distance dependent part of the Casimir energy which be means of $F=-\frac{\pa}{\pa a}E$ is alone responsible for the force. 

The representation \Ref{Eps2} is still to some extend symbolic. In fact, a precise formula is
\be\label{Edet}E=\frac{1}{2\pi} \int_0^\infty d\om  \  \sum_{m=-\infty}^\infty \ln \det \left(\delta_{l,l'}-A_{l,l'} \right),
\ee
where  $\left(\delta_{l,l'}-A_{l,l'}\right) $ is a infinite dimensional matrix with $l,l'\ge|m|$.

Another way to represent the trace in an explicit formula is to expand the logarithm,
\be\label{Esum}E=\frac{1}{2\pi} \int_0^\infty d\om  \ \sum_{s=0}^\infty \frac{-1}{s+1} \sum_{m=-\infty}^\infty 
\sum_{l=|m|}^\infty \sum_{l_1=|m|}^\infty \dots \sum_{l_s=|m|}^\infty 
A_{l,l_1} A_{l_1,l_2}\dots A_{l_s,l}  \ ,
\ee
whereby the convergence of the sum over $s$ is assumed.

Let us mention that the representations  \Ref{Edet} and \Ref{Esum} of the Casimir energy are exact. Their  basic merit  is that they are free of ultraviolet divergences. Hence they can be used also for direct numerical approximation. Examples of that are given in \cite{Bulgac:2005ku} and \cite{Emig:2006uh}.

A further simplification appears because the matrix elements $A_{l,l'}$ can be written explicitly in terms of modified Bessel functions and 3j-symbols. In fact, the formula
\bea\label{Alls}A_{l,l'}&=&\sqrt{\frac{\pi}{2}}\sqrt{\frac{(2l+1)(2l'+1)}{2a\om}}
(-1)^{l+l'}\frac{I_{l'+\frac12}(\om R)}{K_{l+\frac12}(\om R)}
 \\ &&\nn
\cdot\sum_{l''=|l-l'|}^{l+l'}   (-i)^{l''}(2l''+1)K_{l''+\frac12}(2a\om)
\left(\begin{array}{ccc}l''&l&l'\\0&0&0\end{array}\right)
\left(\begin{array}{ccc}l''&l&l'\\0&-m&m\end{array}\right)
\eea
holds. Its derivation is given in the Appendix A. 

Similar formulas as for the sphere can be written for a cylinder in front of a plane. For comparison we consider besides the (3+1)-dimensional also the (2+1)-dimensional case, i.e., a circle in a front of a line on a plane. We denote the energies by $E^{\rm cyl}_{(3+1)}$ and $E^{\rm cyl}_{(2+1)}$ correspondingly.

We start from the formula \Ref{Eom} for the energy which for $E^{\rm cyl}_{(2+1)}$ does not change,
\be\label{Ec20}E^{\rm cyl}_{(2+1)}=\frac{1}{2\pi}\int_{0}^\infty d\om  \   \ tr \ln K_\om \ . 
\ee
For (3+1) dimensions using the translational invariance along the axis of the cylinder it must be substituted by
\be\label{}E^{\rm cyl}_{(3+1)}=\frac{1}{2\pi}\int_{0}^\infty d\om  \ \int_{-\infty}^\infty\frac{d k_z }{2\pi} \ tr \ln K_\gamma
\ee
with $\gamma=\sqrt{\om^2+k_z^2}$ where $k_z$ is the momentum parallel to the axis of the cylinder. Obviously the integration over $k_z$ can be done,
\be\label{Ec30a}E^{\rm cyl}_{(3+1)}=\frac{1}{4\pi}\int_{0}^\infty d\om  \ \om \  \ tr \ln K_\om \ ,
\ee
so that the difference is just the additional factor $\om/2$ in \Ref{Ec30a} as compared to \Ref{Ec20}. 
We note that $E^{\rm cyl}_{(3+1)}$ is the energy per unit length of the cylinder.

Next we have to consider the function $K_\om$ in two dimensions. A basis of functions on a circle is simply $|m>=\exp\{im\varphi\}/\sqrt{2\pi}$ and the free space propagator has the decomposition
\be\label{D2_1s}D^{(2+1)}_\om(x-x')=\sum_ m \mid m\rangle \   d^{(2+1)}_m(r,r')\ \langle m'\mid
\ee
with
\be\label{D2+1c}d^{(2+1)}_m(r,r')=I_m(\om r_<) K_m(\om r_>). 
\ee
In this way we obtain in place of \Ref{Klls}
\be\label{}K_{m,m'}=\delta_{m,m'}d^{(2+1)}_m(R,R)-\langle m\mid D^{(2+1)}_\om(\tilde{x}-x')
\mid m'\rangle \ ,
\ee
where $D^{(2+1)}_\om(\tilde{x}-x')$ is the same as \Ref{Dom3} in a  dimension lower by one,
\be\label{D2+1f}D^{(2+1)}_\om({x}-x')=\int\frac{d^2k}{(2\pi)^2}\ \frac{e^{-ik(x-x')}}{\om^2+k^2}
\ee
and $\tilde{x}=(x,2a-y)$ is the reflected coordinate in the (x,y)-plane.

Separating again the distance independent contribution we represent the energy in the form
\be\label{Ec30}E^{\rm cyl}_{(3+1)}=\frac{1}{4\pi}\int_{0}^\infty d\om  \ \om \  \sum_m \ln \left(\delta_{m,m'}-A_{m,m'}\right)
\ee
with
\be\label{Amms0}A_{m,m'}=\frac{1}{d^{(2+1)}_m(r,r)}\langle m\mid D^{(2+1)}_\om(\tilde{x}-x')
\mid m'\rangle \ .
\ee
%
In the (2+1)-dimensional case we have the same formula except for the factor $\om$ dropped,
\be\label{Ec2}E^{\rm cyl}_{(2+1)}=\frac{1}{2\pi}\int_{0}^\infty d\om  \ \  \sum_m \ln \left(\delta_{m,m'}-A_{m,m'}\right)\ .
\ee

Like in the case of a sphere, the matrix elements $A_{m,m'}$ can be calculated explicitly (see Appendix A) and  expressed in terms of Bessel functions,
\be\label{Amms}A_{m,m'}=\frac{1}{K_m(\om r)}K_{m+m'}(2\om a) I_m(\om r)
\ee
a formula which coincides with the corresponding ones in \cite{Emig:2006uh} and \cite{Bulgac:2005ku}.

\section{Cylinder at small separation from a plane}
In this section we consider a cylinder at small separation from a plane. We start with Dirichlet boundary conditions. Here we consider first the case of (3+1) dimensions, i.e., the usual space. Then we reduce the dimension by one, i.e., we consider the (2+1)-dimensional case which is a circle in front of a line on a plane.
Finally, we consider the case of Neumann boundary conditions. 
  Since in the cylindrical case the polarizations of the electromagnetic field separate into TM modes fulfilling  Dirichlet boundary conditions and TE modes fulfilling Neumann boundary conditions this gives at once the result for the electromagnetic field.

\subsection{Dirichlet case}
Starting point is the formula \Ref{Ec30}. We rescale $\om\to \om /r$ so that the energy is now given by
\be\label{Ec1a} E^{\rm cyl}_{(3+1)}=\frac{1}{4\pi r^2}\int_0^\infty  d\om \ \om \
\tr \ln \left(\delta_{m,m'}-A_{m,m'}(\om)\right)\ ,
\ee
where $A_{m,m'}$ is now given by
\be\label{Ac2}A_{m,m'}(\om)=\frac{1}{K_m(\om)}\ K_{m+m'}(2\om(1+\ep))\ I_{m'}(\om)
\ee
with
\be\label{defep}\ep=\frac{L}{R}
\ee
where $L=a-R$ is the separation between the cylinder and the plane
and we are interested in the small distance behavior, i.e., in $\ep\to0$.

To start with, we expand the logarithm in Eq.\Ref{Ec1a}. Below this turns out to be justified  because in the orders in $\ep$ we are interested in this series converges. The energy is now given by
\be\label{Ec3}E^{\rm cyl}_{(3+1)}=\frac{-1}{4\pi R^2}\sum_{s=1}^\infty\frac{1}{s+1}\int_0^\infty d\om \ \om \ \ \sum_m \sum_{n_1} \dots \sum_{n_{s}} \  {\cal M}
\ee
with 
\be\label{M}{\cal M}=
 A_{m,m+n_1}(\om)A_{m+n_1,m+n_2}(\om)\dots A_{m+n_{s-1},m+n_{s}}(\om)A_{m+n_{s},m}(\om) .
\ee
In  \Ref{Ec3} the summations over $m$ and the $n_i$'s run over the integers.

Now the small distance behavior of the energy comes from the region in the integration resp. sums of large all, $\om$, $m$ and the $n_i$'s.
We use the uniform asymptotic expansion of the Bessel functions and substitute the sums by integrals. 
Further we use the symmetry under $m\to -m$ and obtain in the sense of an asymptotic expansion
\be\label{Ec4}E^{\rm cyl}_{(3+1)}=\frac{-1}{2\pi R^2}\sum_{s=1}^\infty\frac{1}{s+1}\int_0^\infty d\om \ \om \ \ \int_0^\infty dm 
\ 
\int_{-\infty}^\infty dn_{1} \dots \int_{-\infty}^\infty dn_{s} \quad {\cal M},
\ee
where for the $A_{m,m'}(\om)$ in ${\cal M}$ still Eq.\Ref{Ac2} holds, however, with the Bessel functions replaced by their well known uniform asymptotic expansions which are displayed in the Appendix B.
From the expressions in the exponential factors in
\be A_{m,m'}\sim e^{-\tilde{\eta}}\ ,
\ee
which combine into
\be\label{etilde}\tilde{\eta}=(m+m')\ \eta\left(\frac{2\om(1+\ep)}{m+m'}\right)
-m\ \eta\left(\frac{\om}{m}\right)-m' \ \eta\left(\frac{\om}{m'}\right) \ 
\ee
(see Eqs.\Ref{Ac2} and \Ref{ua}),  it follows that for small $\ep$ the dominating contribution into \Ref{Ec4} comes from $\om\sim \ m\sim\frac{1}{\ep}$, $n_i\sim\frac{1}{\sqrt{\ep}}$. Hence we substitute
\bea\label{st2}\om&=&\frac{t\sqrt{1-\tau^2}}{\ep}, \qquad m=\frac{t\tau}{\ep} ,\nn \\
n_i&\to & n_i\ \sqrt{\frac{4t}{\ep}},
\eea
where the factor $\sqrt{4t}$ was introduced for later convenience. With these substitutions the energy becomes
\be\label{Ec5}E^{\rm cyl}_{(3+1)}=\frac{-\ep^{-3}}{2\pi R^2} \sum_{s=0}^\infty \frac{1}{s+1} \int_0^\infty\frac{dt}{t}\ t^3\int_0^1d\tau \  \int_{-\infty}^\infty dn_1\dots  \int_{-\infty}^\infty dn_{s} \ \left(\frac{4t}{\ep}\right)^\frac{s}{2} \ {\cal M}.
\ee
Now we change the notations for the $A_{m,m'}(\om)$,
\be\label{As1}A_{m+n,m+n'}(\om)\to A_{n,n'}^{\rm as}
\ee
indicating the dependence on the variables $n$ and $n'$. Using  the asymptotic expansions  of the Bessel functions entering by means of Eq.\Ref{Ac2},  we obtain the expansion
\be\label{Aas}A_{n,n'}^{\rm as}=\sqrt{\frac{\ep}{4\pi t}} \ \ e^{-{\eta^{\rm as}}} \ \left(1+\sqrt{\ep} \ a_{n,n'}^{(1/2)}(t,\tau) +\ep \ a_{n,n'}^{(1)}(t,\tau) +\dots\right)
\ee
with
\be\label{}{\eta^{\rm as}}=2t+(n-n')^2.
\ee
Details including the explicit form of the functions $ a_{n,n'}^{(1/2)}(t,\tau)$ and  $a_{n,n'}^{(1)}(t,\tau) $ are shown in  Appendix B.

Next we insert \Ref{Aas} for the $A_{m,m'}(\om)$ into $\cal M$ in \Ref{M} and after a reexpansion we obtain 
\bea\label{Mas}{\cal M}&=&\left(\frac{\ep}{4\pi t}\right)^\frac{s+1}{2} \ \ e^{-2(s+1)t-\eta_1}
\left(1+\sqrt{\ep}\ \sum_{i=0}^s  a_{n_i,{n}_{i+1}}^{(1/2)}(t,\tau)
\right.  \\ && \left. +\ep\left(\sum_{0\le i <  j\le s}  a_{n_i,{n}_{i+1}}^{(\frac{1}{2})}(t,\tau)a_{n_j,{n}_{j+1}}^{(\frac{1}{2})}(t,\tau)
 +\sum_{i=0}^s  a_{n_i,{n}_{i+1}}^{(1)}(t,\tau)
 \right)+\dots\right)\nn
\eea
with
\be\label{eta1}\eta_1= \sum_{i=0}^{s}\left(n_i-n_{i+1}\right)^2.
\ee
Here, in order to include all contributions into the sum signs, we have  to put $n_0=n_{s+1}=0$ formally.

The above formulas allow us to to represent the energy $E$, Eq.\Ref{Ec5}, in the form
\be\label{60}E^{\rm cyl}_{(3+1)}=\frac{-1}{2\pi L^2}\sqrt{\frac{R}{L}}\sum_{s=0}^\infty\frac{1}{s+1}
 \int_0^\infty\frac{dt}{t}\ \frac{t^\frac{5}{2} e^{-2(s+1)t}}{\sqrt{4\pi}} \ \int_0^1d\tau \  \int_{-\infty}^\infty \frac{dn_1}{\sqrt{\pi}}\dots  \int_{-\infty}^\infty \frac{dn_{s} }{\sqrt{\pi}} \ \ {\cal M}^{\rm as}
\ee
with
\bea\label{17}{\cal M}^{\rm as}&=&e^{-\eta_1}\left(1+\sqrt{\ep}\ \sum_{i=0}^s  a_{n_i,{n}_{i+1}}^{(1/2)}(t,\tau)
\right.  \\ && \left. +\ep\left(\sum_{0\le i <  j\le s}  a_{n_i,{n}_{i+1}}^{(1/2)}(t,\tau)a_{n_j,{n}_{j+1}}^{(1/2)}(t,\tau)
 +\sum_{i=0}^s  a_{n_i,{n}_{i+1}}^{(1)}(t,\tau)
 \right)+\dots\right)\nn.
\eea
In this representation it is possible to carry out the integrations over $\tau$ and over $t$ and we obtain
\bea\label{Ec6}E^{\rm cyl}_{(3+1)}&=&\frac{-1}{2\pi L^2}\sqrt{\frac{R}{L}}\sum_{s=0}^\infty\frac{1}{s+1}
\int_{-\infty}^\infty \frac{dn_1}{\sqrt{\pi}}\dots  \int_{-\infty}^\infty \frac{dn_{s} }{\sqrt{\pi}} \ \ e^{-\eta_1} \ \  \Bigg\{
\frac{\Gamma(5/2)}{\sqrt{4\pi}(2(s+1))^{5/2}}   \nn  \\ &&   
+\ep \ \ 
\left(
\sum_{0\le i <  j\le s} b(n_i,n_{i+1},n_j,n_{j+1})
+\sum_{i=0}^{s} d(n_i,n_{i+1})+\dots
\right)+\dots\Bigg\}
\eea
with
\bea \label{b}
&&b(n_i,n_{i+1},n_j,n_{j+1}) =
\nn \\ &&
~~~~~~~~~~~~~\int_0^\infty\frac{dt}{t}\ \frac{t^\frac{5}{2} e^{-2(s+1)t}}{\sqrt{4\pi}} \ \int_0^1d\tau \  \
a_{n_i,{n}_{i+1}}^{(1/2)}(t,\tau)a_{n_j,{n}_{j+1}}^{(1/2)}(t,\tau)
\eea
and
\bea \label{d}
d(n_i,n_{i+1}) &=&
\int_0^\infty\frac{dt}{t}\ \frac{t^\frac{5}{2} e^{-2(s+1)t}}{\sqrt{4\pi}} \ \int_0^1d\tau \  \
 a_{n_i,{n}_{i+1}}^{(1)}(t,\tau) \ .
\eea
The quite lengthy explicit formulas for $b$ and $d$ are shown at the end of Appendix B, Eqs. \Ref{bend} and \Ref{dend}.
In Eq.\Ref{Ec6} we dropped the contribution proportional to $\sqrt{\ep}$. It can be seen that it contains only odd powers of the $n_i$'s and vanishes under the integration over the $n_i$'s. 

The next step is to perform the integrations over the $n_i$'s in Eq.\Ref{Ec6}. In the leading order in $\ep$ this is quite simple. We rewrite the quadratic form  $\eta_1$, Eq.\Ref{eta1}, by noticing that obviously the formula
\bea\label{rew}&&n_1^2+\sum_{i=1}^{s-1}\left(n_i-n_{i+1}\right)^2+n_{s}^2\ =\nn
\frac{2}{1}\left(n_1-\frac{2}{1}n_2\right)^2+
\frac{3}{2}\left(n_2-\frac{3}{2}n_3\right)^2+   \dots   \nn \\ &&~~~~~~~~~~~~~~~~~+
\frac{s}{s-1}\left(n_{s-1}-\frac{s-1}{s}n_s\right)^2    +
\frac{s+1}{s} \ n_s^2
\eea
holds.  It  immediately delivers 
\be\label{}\int_{-\infty}^\infty \frac{dn_1}{\sqrt{\pi}}\dots  \int_{-\infty}^\infty \frac{dn_{s} }{\sqrt{\pi}} \ \ e^{-\eta_1} \ = \frac{1}{\sqrt{s+1}}
\ee
and we obtain the energy for a cylinder in front of a plane in leading order in small separation,
\bea\label{Epft}{E^{\rm cyl}_{(3+1)}}^{\rm PFT}&=&\frac{-1}{2\pi L^2}\sqrt{\frac{R}{L}} \ \sum_{s=0}^\infty\frac{\Gamma(5/2)}{\sqrt{4\pi} } \frac{2^{-5/2}}{(s+1)^4}\nn \ ,  \\ 
&=&\frac{-1}{2\pi L^2}\sqrt{\frac{R}{L}} \ \frac{3\zeta(4)}{32\sqrt{2}} =
-\frac{\pi^3}{1920\sqrt{2}}\frac{1}{L^2}\sqrt{\frac{R}{L}} \ ,
\eea
which coincides with the known result from applying the proximity force theorem. Note that in order to restore dimensions it has to be multiplied with $\hbar$ and $c$. Also note that this is the energy per unit length of the cylinder. 

Now we turn to the remaining terms in Eq.\Ref{Ec6}. Of course, all integrations there over the $n_i$'s  are Gaussian. But the combinatorics is somewhat involved so that we delegate these calculations into Appendix C. The result is enjoyable simple,
\bea
\label{longint}
 &&\int_{-\infty}^\infty \frac{dn_1}{\sqrt{\pi}}\dots  \int_{-\infty}^\infty \frac{dn_{s} }{\sqrt{\pi}} \ 
\left(\sum_{0\le i <  j\le s}  b(n_1,n_{i+1},n_j,n_{j+1})
 +\sum_{i=0}^s  d(n_i,{n}_{i+1})
 \right)
 \nn \\ &&
 =\frac{7}{384\sqrt{2}}\frac{1}{(1+s)^3}.
 \eea
The remaining sum over $s$ in \Ref{Ec6} is the same as in the leading order and finally for small distance $L$ the energy becomes     
\be\label{Efin3}
E^{\rm cyl}_{(3+1)}=\frac{-1}{2\pi L^2}\sqrt{\frac{R}{L}}
 \ \frac{3\zeta(4)}{32\sqrt{2}}
 \left\{
1
+ \  \frac{7}{36} \frac{L}{R}
+O\left(\left(\frac{L}{R}\right)^2\right)
\right\} ,
\ee
a formula, which represents the first correction to the proximity force theorem.

Using the method shown above it is easy to repeat the calculation for (2+1) dimensions. 
The first change is in the substitution \Ref{st2} where it is now meaningful not to use the variable $\tau$ but to define
\bea\label{st3}\om&=&\frac{t\sin\varphi}{\ep}, \qquad m=\frac{t\cos\varphi}{\ep} ,\nn \\
n_i&\to & n_i\ \sqrt{\frac{4t}{\ep}},
\eea
with $\varphi\in [0,\frac{\pi}{2}]$. Next we mention the changed formula \Ref{60}
\be\label{62}E^{\rm cyl}_{(2+1)}=\frac{-1}{2 L}\sqrt{\frac{R}{L}}\sum_{s=0}^\infty\frac{1}{s+1}
 \int_0^\infty\frac{dt}{t}\ \frac{t^\frac{3}{2} e^{-2(s+1)t}}{\sqrt{4\pi}} \ \int_0^{\pi/2}\frac{d\varphi}{\pi/2} \  \int_{-\infty}^\infty \frac{dn_1}{\sqrt{\pi}}\dots  \int_{-\infty}^\infty \frac{dn_{s} }{\sqrt{\pi}} \ \ {\cal M}^{\rm as}\ ,
\ee
where besides the prefactor only the power of $t$ changed and the integration over $\tau$ is changed for that over $\varphi$. The remaining steps go in complete parallel to the (3+1)-dimensional case and the result is
\be\label{Efin2}
E^{\rm cyl}_{(2+1)}=\frac{-1}{2 L}\sqrt{\frac{R}{L}}
 \ \frac{\zeta(3)}{8\sqrt{2}}
 \left\{
1
+
 \  \frac{1
}{4} \frac{L}{R}
+O\left(\left(\frac{L}{R}\right)^2\right)
\right\} \ .
\ee
\subsection{Neumann Case}
Let us consider the changes which come in for Neumann boundary conditions first in the general formulas of section 2. Basically it is sufficient to change in formula \Ref{H} the delta function for its normal (to the surface $S$) derivative, $H\to \pa_{n_x}\delta(x-f(z))$. As a consequence, the definition of $K(z,z')$, \Ref{K}, changes for
\be\label{KN}K(z,z')=\pa_{n_x}\pa_{n_{x'}}D(f(x),f(x')),
\ee
where the normal derivatives must be taken before putting the argument onto the surface. The next changes is in formula \Ref{DD0} because on the plane at $z=a$ we have to take Neumann conditions too. It amounts simply in a different sign,
\be\label{NN0} D(x,x')\to D_N(x,x')=D(x-x')+D(\tilde{x}-x')\ .
\ee
As next changes we mention formula \Ref{D2+1c} where the Bessel functions must be substituted by their derivatives,
\be\label{D2+1cN}d^{(2+1)}_m(r,r')=I'_m(\om r_<) K'_m(\om r_>). 
\ee
and equation \Ref{Ec30} which looks now
\be\label{Ec30aN}E^{\rm cyl, N}_{(3+1)}=\frac{1}{4\pi}\int_{0}^\infty d\om  \ \om \  \sum_m \ln \left(\delta_{m,m'}+A_{m,m'}\right)
\ee
with
\be\label{AmmsN}A_{m,m'}=\frac{1}{K'_m(\om R)}K_{m+m'}(2\om a) I'_m(\om R)
\ee
Now we trace the changes in the asymptotic expansion for small separations. 
First we consider the asymptotic expansion of $A_{m,m'}$, \Ref{AmmsN}, using the asymptotic expansions of the derivatives of the Bessel functions,
\be\label{uader}\left. {I'_m(mz)\atop K'_m(mz)}\right\} =\pm \frac{\pi^{\mp\frac12}}{\sqrt{2 m }}\frac{(1+z^2)^{1/4}}{z}\ e^{\pm m\eta(z))}
\left(1\pm \frac{v_1(t)}{m}+\dots\right)
\ee
with $v_1(t)=(-9t+7t^3)/24$. The first observation is that the minus sign in the expansion of $K'_m(mz)$ just compensates the sign change in \Ref{Ec30aN}. The second remark is that in the expansion of the factors in front of the exponential, \Ref{A4} in the Dirichlet case, just $n$ and $n'$ change their places which does not affect the expression under the trace in \Ref{Ec30aN}. The final remark is that the only change comes from the Debye polynomial $v_1(t)$ which appears in two places instead of $u_1(t)$. The further calculation runs then in exactly the same way as in the Dirichlet case, but the result is different, namely for  the formula which corresponds to \Ref{longint} 
we obtain 
\bea
\label{longintN}
 &&\int_{-\infty}^\infty \frac{dn_1}{\sqrt{\pi}}\dots  \int_{-\infty}^\infty \frac{dn_{s} }{\sqrt{\pi}} \ 
\left(\sum_{0\le i <  j\le s}  b(n_1,n_{i+1},n_j,n_{j+1})
 +\sum_{i=0}^s  d(n_i,{n}_{i+1})
 \right)
 \nn \\ &&
 =\frac{7}{384\sqrt{2}}\frac{1}{(1+s)^3}-\frac{1}{12\sqrt{2}(s+1)} \ ,
 \eea
\be\label{}
\ee
i.e., we get a contribution which is the same as in the Dirichlet case and an additional contribution. Inserted into the energy this gives
\be\label{Efin3N}
E^{\rm cyl, N}_{(3+1)}=\frac{-1}{2\pi L^2}\sqrt{\frac{R}{L}}
 \ \frac{3\zeta(4)}{32\sqrt{2}}
 \left\{
1
+ \ \left( \frac{7}{36} -\frac89\frac{\zeta(2)}{\zeta(4)}\right)\frac{L}{R}
+O\left(\left(\frac{L}{R}\right)^2\right)
\right\} ,
\ee
a formula, which represents the first correction to the proximity force theorem for Neumann boundary conditions.

We note that the mentioned compensation of the signs ensures that the leading order contribution, i.e., the proximity force theorem, gives the same answer for both types of boundary conditions as expected. Less obvious is the behavior of the Neumann contribution in what it has in  addition to the Dirichlet case (which equals the first term in the r.h.s. of Eq. \Ref{longintN}). This is the second term in the r.h.s. of Eq. \Ref{longintN}, which results in a  much less convergent sum over $s$. In the (3+1) dimensional case the sum is still convergent, but in the (2+1) dimensional case not. Carrying out the corresponding calculation we obtain 
\be\label{}\frac{1}{32\sqrt{2}(s+1)^2}-\frac{1}{4\sqrt{2}}
\ee
in place of the r.h.s. of Eq.\Ref{longintN}. Now the sum over $s$ is logarithmic divergent. In fact this implies that the correction to the proximity force theorem in (2+1) dimensions for Neumann  boundary conditions is proportional to $\frac{L}{R}\ln(L/R)$. 

Finally we collect the energies for Dirichlet and Neumann boundary conditions together to the first correction to the proximity force theorem for the electromagnetic fields,

\bea\label{Elmagn}
E^{\rm cyl, electromgn.}_{(3+1)}&=&\frac{-1}{2\pi L^2}\sqrt{\frac{R}{L}}
 \ \frac{3\zeta(4)}{16\sqrt{2}}
 \left\{
1
+ \ \left( \frac{7}{36} -\frac49\frac{\zeta(2)}{\zeta(4)}\right)\frac{L}{R}
+O\left(\left(\frac{L}{R}\right)^2\right)
\right\} ,
\nn \\ &=&
-\frac{1}{L^2}\sqrt{\frac{R}{L}}\frac{\pi^3}{960\sqrt{2}} 
\left\{
1
+ \ \underbrace{\left( \frac{7}{36} -\frac{20}{3\pi^2}\right)}\frac{L}{R}
+O\left(\left(\frac{L}{R}\right)^2\right)
\right\} .
\nn \\ && ~~~~~~~~~~~~~~~~~~~~~~~~~~~~~~~~~~~~{=-0.48103}
\eea
%

\section{Sphere and cylinder at large separation from a plane}
The limiting case of large separation between a sphere and a cylinder and a plane is known and we display it here for completeness. 
For a sphere the starting point is Eq.\Ref{Esum} with $A_{l,l'}$ given by Eq.\Ref{Alls}. The main contribution comes from small $\om$ and small orbital momenta. We substitute 
\be\label{sul}\om\to\frac{\om}{2a}
\ee
and expand $A_{l,l'}$ for small $R/a$. simply by expanding the two Bessel functions in \Ref{Alls} which depend on $r$. The leading order is
\be A_{l,l'}=\delta_{l,0}\delta_{l',0}\ \frac{R}{2a} \ e^{-\om}+\dots \ .
\ee
Inserted into $E$, \Ref{Esum}, this gives immediately the large distance limit 
\be
E=-\frac{r}{8\pi a^2}\left(1+\frac34 \ \frac{R}{a} + O\left(\left(\frac{R}{a}\right)^2\right)\right)\ ,
\ee
where we included the first correction too (note $a=L+R$). This result agrees with \cite{Bulgac:2005ku}.

The corresponding procedure for a cylinder is similar. We start from Eq.\Ref{Ec2} or from Eq.\Ref{Ec3} in dependence on the dimension and with $A_{m,m'}$ given by Eq.\Ref{Amms}. Again, the dominating contribution comes from small momenta. We do the substitution \Ref{sul} and expand $A_{m,m'}$. Here, however, a peculiarity shows up because of the logarithmic behavior of the Bessel function $K_m(\om\frac{r}{2a})$ in the denominator for $m=0$.  The expansion is
\be A_{m,m'}=\frac{\delta_{m,0}\delta_{m',0} \ K_0(\om)}{-\gamma-\ln(\om R/4a)}+O\left(\left(\frac{R}{a}\right)^2\right) \ ,
\ee
where $\gamma$ is Euler's constant. Inserting into \Ref{Esum} and integrating over $\om$ we get
\be
E^{\rm cyl}_{(2+1)}=\frac{\pi}{8a} \ 
\frac{1}{\ln(R/2a)}
\left(1-\frac{2\ln 2}{\ln(R/2a)}+O\left(\left(\frac{1}{\ln(R/2a)}\right)^2\right)
\right)+O\left(\left(\frac{R}{a}\right)^2\right)
\ee
and
\be
E^{\rm cyl}_{(3+1)}=\frac{1}{8\pi a^2} \ 
\frac{1}{\ln(R/2a)}
\left(1-\frac{\pi^2/12}{(\ln(R/2a))^2}+O\left(\left(\frac{1}{\ln(R/2a)}\right)^3\right)
\right)+O\left(\left(\frac{R}{a}\right)^2\right).
\ee
In the (3+1)-dimensional case the next-to-leading order is zero. 

We note that the logarithmic behavior of the large distance limit is related to the corresponding behavior of the Greens function in two dimensions. Because the logarithmic behavior is present in $E^{\rm cyl}_{(2+1)}$, i.e., for a circle on a plane, too it seems unlikely that it is related to the length of the cylinder as discussed for the three dimensional case in \cite{Emig:2006uh}.

\section{Conclusions}
In the foregoing sections we showed the derivation of the expression for the Casimir energy in terms of the determinant of the projection $K(z,z')$ of the free space propagator on the boundary surface $S$ which was known from \cite{Bordag:1985zk} and \cite{Robaschik:1986vj}
and independently rederived and applied to the calculation at large distance in 
\cite{Bulgac:2005ku} and \cite{Emig:2006uh} which can be considered as a significant progress because it opened the way for direct numerical calculation of the Casimir force for non simple geometries. In the present paper we showed that also the short distance behavior can be calculated by this method and derived the first corrections to the force proximity theorem including the electromagnetic case. In (3+1) dimensions the corrections are  power like, in (2+1) dimensions the corrections contain a logarithm for Neumann conditions. 

In general, corrections to the limiting behavior for both, large and small distance, can be calculated as expansion in the corresponding small parameter. These expansions are asymptotic ones. For the small distance behavior this is clear from section 4 where we substituted the sums over the orbital momenta of the photon propagator by the corresponding integrals. Obviously, exponentially small contributions are neglected in this way. For the long distance expansion this follows from the denominators in the corresponding matrix elements, \Ref{Alls} and \Ref{Amms}. Consider the energy as a function of the expansion parameter $\delta=r/2a$ (after the rescaling\Ref{sul}) in the complex plane of $\delta$. 
When moving $\delta$ around the origin the pole of the integrand crosses at some time the path of the integration over $\om$ delivering an additional contribution. Hence, the energy as a function of $\delta$ has a cut starting from $\delta=0$. 

The new method for the Casimir force for non simple geometry which so far gave the possibility to calculate the large and the small distance behavior, has the potential in it for the calculation of the Casimir force at all distances with desired precision. Also, it obviously can be generalized to much more boundary conditions than Dirichlet or Neumann ones. In the present paper we considered primarily Dirichlet conditions  in order to keep the representation as simple as possible. The generalizations to Neumann are obvious, one has to change the sign in \Ref{DD0} and to add a derivative to $H(x,z)$, Eq.\Ref{H} as done, for example, in \cite{Emig:2006uh}. In the path integral approach this was considered from a more general point of view in  \cite{Grosche:1994uv}. 

The short distance expansion in section 4 sheds light also on the quasi classic, optical path and similar approximations. Indeed, the dominating contributions come solely from large momentum, however in an expansion over the number of reflections, like the expansion of the logarithm in section 4 shows, all terms contribute. Therefore, in order to get some approximation, say in the first order of the small parameter, all reflections must be included. 

\section*{Acknowledgements}
This work was supported by the research funding from the EC's Sixth Framework Programme within the STRP project "PARNASS" (NMP4-CT-2005-01707).\\
{\it Note added}. -— Since the submission of this paper, Gies
and Klingmu\"uller \cite{Gies:2006cq} confirmed the correction factor $7/36$
in Eq. (69) for Dirichlet boundary conditions in the
cylinder-plane geometry by numerical world line methods.

\setcounter{equation}{0}\renewcommand{\theequation}{A.\arabic{equation}}
\section*{Appendix A}
In this appendix we show the calculation of the matrix elements \Ref{Amms0} and \Ref{Alls0} in section 3. We start with the cylindrical case. Using   \Ref{D2+1f} we represent the matrix element in \Ref{Amms0} in the form
\bea\label{W1}W_{m,m'}&\equiv&\langle m\mid D^{(2+1)}_\om(\tilde{x}-x')
\mid m'\rangle \nn \\ &=&
\int\frac{d^2k}{(2\pi)^2}\ \frac{e^{2iak_2}}{\om^2+k^2} \ f_m(\tilde{k}) \ f_m^*(k)
\eea
with
\be\label{fm}f_m(k)=\int_0^{2\pi}\frac{d\varphi}{\sqrt{2\pi}} \ \ e^{-im\varphi+ikx} \ . 
\ee
Here we used $\tilde{x}=(x_1,2a-x_2)$ and defined $\tilde{k}=(k_1,-k_2)$ in order to rewrite $k(\tilde{x}-x')=2ak_2+\tilde{k}x -kx'$. Using polar coordinates in the $k$-plane, $k_1=k\sin\varphi_k$, $k_2=k\cos\varphi_k$, and an integral representation 
\be\label{irep}\int_0^{2\pi}\frac{d\varphi}{{2\pi}}\ e^{-im\varphi+iz \sin\varphi} = J_m(z)
\ee
of the Bessel function $J_m(z)$ for the functions $f_m(k)$ entering \Ref{W1} we obtain
\be\label{}W_{m,m'}= \int\frac{d^2k}{2\pi}\ \frac{e^{2iak_2}}{\om^2+k^2}\ J_m(kr)\ J_{m'}(kr)\ e^{-i(m+m')\varphi_k}.
\ee
The formula \Ref{irep} can also be used for the integration over the angle $\varphi_k$,
\be\label{}W_{m,m'}= \int_0^\infty\frac{dk\ k}{\om^2+k^2}\ J_m(kr)\ J_{m'}(kr)\ J_{m+m'}(2ka).
\ee
The last integration can be done by extending it over the whole $k$-axis and closing the contour picking up the pole in $k=i\om$. With the relations $J_m(z)=\frac12\left(H^{(1)}_m(z)+H^{(2)}_m(z)\right)$, $H^{(1)}_m(-z)=(-1)^{m+1}H^{(2)}_m(z)$, $J_m(-z)=(-1)^{m}J_m(z)$ one obtains
\be\label{}W_{m,m'}=\frac{i\pi}{2} \ J_m(i\om r)\ J_{m'}(i\om r) \ H^{(1)}_{m+m'}(2i\om a)\ ,
\ee
or in terms of the modified Bessel functions,
\be\label{}W_{m,m'}=   I_m(\om r)\ I_{m'}(\om r)\ K_{m+m'}(2\om a)\ ,
\ee
which after division by $d_m^{(2+1)}(r,r)$, \Ref{D2+1c}, gives \Ref{Amms}.

Now we turn to the spherical case and consider the matrix elements in \Ref{Alls0}. Using \Ref{Dom3} we represent them in the form
\bea\label{}D_{l,l'}&\equiv&\langle lm\mid D_{\om}(x-\tilde{x})\mid l'm \rangle
\nn \\ &=&
\int\frac{d^3k}{(2\pi)^3}\ \frac{e^{-2iak_3}}{\om^2+k^2} \ f^*_{lm}(k) \ f_{l'm}(\tilde{k})
\eea
with $\tilde{k}=(k_1,k_2,-k_3)$ and
\bea\label{}f_{lm}(k)&\equiv&e^{-ikx}\mid lm \rangle
\nn \\ &=& \int_0^\pi d \theta\sin\theta\int_0^{2\pi}d\varphi \ e^{-ikx} \ Y_{lm}(\theta,\varphi)
\nn \\ &=&
4\pi (-i)^lj_l(kr)Y_{lm}(\theta_k,\varphi_k)\ ,
\eea
whereby in the last line the decomposition of a plane wave into spherical waves,
\be\label{sphp}e^{ikz}=4\pi\sum_{lm}i^lj_l(kr) Y^*_{lm}(\theta_z,\varphi_z) Y_{lm}(\theta_k,\varphi_k)\ ,
\ee
was used. Here, $j_l(z)$ are the spherical Bessel functions.
Writing the integration over $k$ in spherical coordinates and using again \Ref{sphp} an integral over a product of three hypergeometric functions appears which can be written in terms of 3j-symbols,
\bea\label{}
&&
\int_0^\pi d \theta\sin\theta\int_0^{2\pi}d\varphi \
Y_{lm}(\theta,\varphi)Y_{l'm'}(\theta,\varphi)Y_{l''m''}(\theta,\varphi)
\nn \\ &&=
\sqrt{\frac{(2l+1)(2l'+1)(2l''+1)}{4\pi}}
\left(\begin{array}{ccc}l&l'&l''\\0&0&0\end{array}\right)
\left(\begin{array}{ccc}l&l'&l''\\m&m'&m''\end{array}\right)
\eea
and we arrive at
\bea\label{}D_{l,l'}&=&\frac{2}{\pi}\sqrt{(2l+1)(2l'+1)}
\int_0^\infty\frac{dk\ k^2}{\om^2+k^2}
j_l(kr)j_{l'}(kr)
 \\ &&\nn   \cdot
\sum_{l''=|l-l'|}^{l+l'} i^{l+l'-l''}
(2l''+1)j_{l''}(2ak)
\left(\begin{array}{ccc}l''&l&l'\\0&0&0\end{array}\right)
\left(\begin{array}{ccc}l''&l&l'\\0&-m&m\end{array}\right).
\eea
Using $j_l(z)=\frac12(h^{(1)}_l(z)+h^{(2)}_l(z))$, the integration can be extended to the whole $k$-axis and closed picking up the pole in $k=i\om$. The result can be inserted into \Ref{Alls0} and we obtain
\bea\label{}A_{l,l'}&=&\sqrt{(2l+1)(2l'+1)}\frac{j_{l'}(ir\om)}{h^{(1)}_l(ir\om)}
 \\ &&\nn   \cdot
\sum_{l''=|l-l'|}^{l+l'} i^{l+l'-l''}
(2l''+1)
h^{(1)}_{l''}(2ia \om )
\left(\begin{array}{ccc}l''&l&l'\\0&0&0\end{array}\right)
\left(\begin{array}{ccc}l''&l&l'\\0&-m&m\end{array}\right)
\eea
which coincides with the corresponding formula in \cite{Bulgac:2005ku}. Passing now to the modified Bessel functions and dividing by $d_l(r,r)$, \Ref{dl25}, we obtain just formula \Ref{Alls}.

\setcounter{equation}{0}\renewcommand{\theequation}{B.\arabic{equation}}
\section*{Appendix B}
In this appendix we calculate the uniform asymptotic expansion of $A_{m,m'}(\om)$, Eq.\Ref{Ac2}, using \Ref{ua} and obtain $A_{m,m'}^{\rm as}$, Eq.\Ref{Aas}, and the coefficients $a_{n,n'}^{(1/2)}(t,\tau)$ and $a_{n,n'}^{(1)}(t,\tau)$ therein. 

We remind the uniform asymptotic expansion of the Bessel functions,
\be\label{ua}\left. {I_m(mz)\atop K_m(mz)}\right\} =\frac{\pi^{\mp\frac12}}{\sqrt{2 m }}\frac{\exp(\pm m\eta(z))}{(1+z^2)^{1/4}}
\left(1\pm \frac{u_1(t)}{m}+\dots\right)
\ee
with $\eta(z)=\sqrt{1+z^2}+\ln(z/(1+\sqrt{1+z^2}))$ and  $u_1(t)=(3t-5t^3)/24$ with $t=1/\sqrt{1+z^2}$, the first of the Debye polynomials.
Since we intend to calculate the first next-to-leading order only  the restriction to $u_1(t)$  is sufficient.

We start with the exponential factor and do the substitution \Ref{st2} there. Up to order $\ep$ we get
\bea\label{A1}\tilde{\eta}&=&2t+(n-n')^2
+\sqrt{\frac{\ep}{t}} \ \tau  \ \left(2t-(n - {n'})^2 \right) (n + {n'})  \\ &&+ 
 \ep \ \left(- \tau^2 
 + \frac{1 - \tau^2}{t}  (n + {n'})^2
 - \frac{1 - 3 \tau^2}{12t^2} ((n - {n'})^4 + 6 (n^2- {n'}^2)^2)  \right)  .\nn
\eea
The expansion of the exponential is
\be\label{A2}e^{-\tilde{\eta}}=e^{-\eta_0}\left(1+\sqrt{\ep} \ \eta^{(1/2)}_{n,{n'}} +  \ep \ \eta^{(1)}_{n,n'} +O)\ep^2)\right)
\ee
with
\be\label{A3halbe} \eta^{(1/2)}_{n,{n'}}=-\frac{1}{\sqrt{t}} \ \tau  \ \left(2t-(n - {n'})^2 \right) (n + {n'}) .
\ee
and
\be\label{A3} \eta^{(1)}_{n,{n'}}=-t\ \ \tau^2+(1-\tau^2)(n+n')^2-\frac{1-3\tau^2}{12t}(n-n')^2(7n^2+10n n'+7{n'}^2).
\ee
Next we consider the factors in front of the exponential in $A^{\rm as}_{n,n'}$. From \Ref{Ac2} and \Ref{ua} they collect into
\be\label{A4}C_{m,m'}=\frac{1}{\sqrt{2\pi}} \ \left(\frac{m}{m'}\frac{1}{m+m'}\right)^\frac{1}{2} \left(\frac{1+z_1^2}{1+z_2^2}\frac{1}{1+z_3^2}\right)^\frac{1}{4}
\ee
with $z_1=\frac{\om}{m}$, $z_2=\frac{\om}{m'}$ and $z_1=\frac{2\om (1+\ep)}{m+m'}$. Here, again, we do the substitution \Ref{st2} and the expansion for small $\ep$ gives
\be\label{A5}C_{m,m'}\to c_{m,m'}=\sqrt{\frac{\ep}{4\pi t}}\left(1+\sqrt{\ep} \ c^{(1/2)}_{n,n}+\ep \ c^{(1)}_{n,n'}+\dots\right)
\ee
with 
\be\label{A6a} c^{(1/2)}_{n,{n'}}=\frac{ \tau}{2\sqrt{t}} \ (n-3n')
\ee
and
\be\label{A6b} c^{(1)}_{n,{n'}}=\frac{-1+\tau^2}{2}+
\frac{1}{8t} \ \left(6n^2-4nn'-10{n'}^2+(-11n^2+2nn'+29{n'}^2)\ \tau^2\right).
\ee
We note hat these contributions are not symmetric under $n\leftrightarrow n'$.

Finally we have to consider the contributions from the Debye polynomials. To the given order we note
\be\label{A7}D=\frac{1+\frac{u_1(t_1)}{m}+\dots}{1-\frac{u_1(t_2)}{m'}+\dots}\left(1-\frac{u_1(t_3)}{m+m'}+\dots\right)
\ee
with $t_i=\sqrt{1+z_i^2}$.
 Doing here the substitution \Ref{st2} we obtain 
\be\label{A8}D\to 1+\ep \ d^{(1)}+\dots
\ee
with %
\be\label{A9} d^{(1)}=\frac{3-5\tau^2}{16t} \ .
\ee
Taking \Ref{A2}, \Ref{A5} and \Ref{A8} together and reexpanding again we obtain finally for $a_{n,n'}^{(1/2)}(t,\tau)$ and $a_{n,n'}^{(1)}(t,\tau)$  in Eq.\Ref{Aas}
\bea\label{}a_{n,n'}^{(1/2)}(t,\tau)
&=&\frac{\left(n-3 \text{{n'}}+2 (n+\text{{n'}}) \left(n^2-2 \text{{n'}} n+\text{{n'}}^2-2 t\right)\right) \tau }{2
   \sqrt{t}},
\nn \\ 
a_{n,n'}^{(1)}(t,\tau)&=&
\frac{(n-3 \text{{n'}}) (n+\text{{n'}}) \left(n^2-2 \text{{n'}} n+\text{{n'}}^2-2
   t\right) \tau ^2}{2 t}+\frac{3-5 \tau ^2}{16 t}
\nn \\ &&
   +\frac{1}{8 t} \bigg[ \left(6-11 \tau
   ^2\right) n^2+2 \text{{n'}} \left(\tau ^2-2\right) n+4 t \left(\tau
   ^2-1\right)\nn \\ &&~~~~~~~+\text{{n'}}^2 \left(29 \tau ^2-10\right)\bigg]
\nn \\ &&
   +\frac{1}{12 t} \ \bigg[-7 \left(3
   \tau ^2-1\right) n^4+4 \text{{n'}} \left(3 \tau ^2-1\right) n^3
\nn \\ &&   ~~~~~~~
   +6 \left(\left(3
   \tau ^2-1\right) \text{{n'}}^2+2 t \left(\tau ^2-1\right)\right) n^2
\nn \\ &&~~~~~~~
   +4
   n\text{{n'}} \left(\left(3 \tau ^2-1\right) \text{{n'}}^2+6 t \left(\tau
   ^2-1\right)\right) 
\nn \\ &&~~~~~~~
+6 (n+\text{{n'}})^2 \left(n^2-2 \text{{n'}} n+\text{{n'}}^2-2
   t\right)^2 \tau ^2
   \nn \\ &&~~~~~~~
   +12 t^2 \tau ^2+12 \text{{n'}}^2 t \left(\tau ^2-1\right)-7
   \text{{n'}}^4 \left(3 \tau ^2-1\right)\bigg] \ .
\eea

At the end of this appendix we display the explicit forms of the functions $b$, \Ref{b}, and $d$, \Ref{d}, introduced in Eq.\Ref{Ec6}: 
\bea\label{bend}
&&b(x,y,z,w)=
\frac{1}{384 \sqrt{2} (s+1)^{7/2}}
\Big[ 
4 (s+1) \left(4 (s+1) x^3-4 (s+1) y x^2
\right. \nn \\ && \left.
+2 \left(-2 (s+1) y^2+s-2\right) x+y \left(4 (s+1) y^2-6 (s+1)-6\right)\right) w^3
\nn \\ &&
+4 (s+1) \left(-4 (s+1) x^3+4 (s+1) y
   x^2+\left(4 (s+1) y^2-2 (s+1)+6\right) x
\right.   \nn \\ && \left.
+y \left(-4 (s+1) y^2+6 (s+1)+6\right)\right) z w^2+\left(-4 (s+1) \left(2 (s+1) \left(2 z^2+3\right)+6\right) x^3
\right. \nn \\ &&  \left.
+4 (s+1) y
   \left(2 (s+1) \left(2 z^2+3\right)+6\right) x^2+\left(4 \left(2 y^2-1\right) \left(2 z^2+3\right) (s+1)^2
\right. \right. \nn \\ &&  \left. \left.
+24 \left(y^2+z^2+1\right) (s+1)+60\right) x+y \left(-4
   \left(2 y^2-3\right) \left(2 z^2+3\right) (s+1)^2
\right. \right. \nn \\ &&  \left. \left.
   -24 \left(y^2-z^2-3\right) (s+1)+60\right)\right) w+z \left(8 (s+1) \left(2 s z^2+2 z^2+s-2\right) x^3
\right. \nn \\ &&  \left. 
-8 (s+1) y
   \left(2 s z^2+2 z^2+s-2\right) x^2+\left(-4 \left(2 y^2-1\right) \left(2 z^2+1\right) (s+1)^2
\right. \right. \nn \\ &&  \left. \left.
+24 \left(y^2-z^2-1\right) (s+1)+60\right) x+y \left(4 \left(2
   y^2-3\right) \left(2 z^2+1\right) (s+1)^2
\right. \right. \nn \\ &&  \left. \left.  
   -24 \left(y^2+z^2-1\right) (s+1)+60\right)\right)
\Big] \ ,
 \\ 
&&\label{dend}
d(x,y)=\frac{1}{384 \sqrt{2} (s+1)^{7/2}}\Big[ 8 (s+1)^2 x^6
\nn \\ &&
-16 (s+1)^2 y x^5-8 (s+1) \left((s+1) y^2-s+2\right) x^4
\nn \\ &&
+32 (s+1)^2 y \left(y^2-1\right) x^3-2 \left(4 (s+1)^2 y^4-8 (s+1) (s+4) y^2
\right. \nn \\ && \left.
+(4-7 s) s-4\right)
   x^2+4 y \left(-4 (s+1)^2 y^4+8 (s+1)^2 y^2-s (5 s+16)+4\right) x
\nn \\ &&
+8 (s+1)^2 y^6-24 (s+1) (s+2) y^4+(40-2 (s-4) s) y^2+4 (s-1) s+7 \Big]  .
\eea
 
\setcounter{equation}{0}\renewcommand{\theequation}{C.\arabic{equation}}
\section*{Appendix C}
In this appendix we calculate the integrals over the $n_i$'s in Eq.\Ref{Ec6}. As mentioned in the text all these integrations are Gaussian and we are left with the corresponding combinatorics. 

We note that a number of these integrations can be carried out independently of the 
functions $b(n_i,n_{i+1},n_j,n_{j+1})$ and $d(n_i,n_{i+1})$  entering \Ref{Ec6}. In the following we divide the whole expression into a number of pieces and in each we first perform these integrations. The remaining ones, one to four in number, depend on the details of these functions. They have been performed machined. 

We start with the part involving $d(n_i,n_{i+1})$, \Ref{d}, in \Ref{Ec6}. It is easier because it contains only one sum. First we separate the first and the last contributions to this sum,
\be
\label{B1}\sum_{i=0}^s d(n_i,n_{i+1})=d(0,n_1)+d(n_s,0)+\sum_{i=1}^{s-1} d(n_i,n_{i+1}).
\ee
Renaming the $n_i$'s by numbering them in the reverse order in the first term in the r.h.s. we have
\bea
\label{B2}S_1&\equiv& \int_{-\infty}^\infty \frac{dn_1}{\sqrt{\pi}}\dots  \int_{-\infty}^\infty \frac{dn_{s} }{\sqrt{\pi}} 
\left( d(0,n_1)+d(n_s,0)\right) \ e^{-\eta_1}
\nn \\ &=&
\int_{-\infty}^\infty \frac{dn_1}{\sqrt{\pi}}\dots  \int_{-\infty}^\infty \frac{dn_{s} }{\sqrt{\pi}} 
\left( d(0,n_s)+d(n_s,0)\right) \ e^{-\eta_1}.
\eea
Now we apply formula \Ref{rew}. However, in order to avoid confusion with the notations we rewrite it,
\bea
\label{B3}&&m_1^2+\sum_{i=1}^{s-1}\left(m_i-m_{i+1}\right)^2+m_{u}^2\ = 
\frac{2}{1}\left(m_1-\frac{2}{1}m_2\right)^2+
\frac{3}{2}\left(m_2-\frac{3}{2}m_3\right)^2+   \dots  \nn  \\ && \dots +
\frac{u}{u-1}\left(m_{u-1}-\frac{u-1}{u}m_u\right)^2    +
\frac{u+1}{u} \ m_u^2.
\eea
With $u=s$, $m_1=n_1$, ..., $m_{u-1}=n_{s-1}$  the first ($s-1$) integrations can be carried out, one by one. The first is that over $n_1$, which must be shifted, $n_1\to n_1+\frac{2}{1}n_2$ and gives just $\sqrt{\frac{1}{2}}\sqrt{\pi}$. The second integrations needs the shift $n_2\to n_2+\frac{3}{2}n_3$  and gives $\sqrt{\frac{2}{3}}\sqrt{\pi}$ and so on. The shifts of the variables which do not enter the functions $d$ (and, below, $b$) are obvious and we do not mention them in the following. Finally, renaming  $n_s=x$ we obtain
\be
\label{S10}S_1=\frac{1}{\sqrt{s}} \ \int_{-\infty}^\infty \frac{dx }{\sqrt{\pi}} \
\left( d(0,x)+d(x,0)\right) \ exp\left(-\frac{s+1}{s} x^2\right).
\ee
The last integration was carried out machined (like the similar ones below) and the result reads  
\be
\label{}S_1=\frac{1}{12 \sqrt{2} (s+1)^2}-\frac{3}{8 \sqrt{2} (s+1)^3}+\frac{37}{64 \sqrt{2} (s+1)^4}-\frac{1}{4 \sqrt{2} (s+1)^5}.
\ee

Next we calculate the contribution from the sum over $i$ in \Ref{B1}. To this end we represent $\eta_1$ in the form
\bea
\label{}\eta_1&=&n_1^2+(n_1-n_2)^2+\dots+(n_{i-1}-n_i)^2+n_i^2
\nn \\ &&
-n_i^2+(n_{i}-n_{i+1})^2-n_{i+1}^2
\nn \\ &&
+n_{i+1}^2+(n_{i+1}-n_{i+2})^2+\dots+n_s^2
\eea
and apply \Ref{B3} to $n_1$, ..., $n_i$ with $u=i$ and  $m_1=n_1$, ..., $m_u=n_{i}$ and to 
$n_{i+1}$, ..., $n_s$ with $u=s-i$ and  $m_1=n_s$, ..., $m_u=n_{i+1}$ so that it takes the form 
\bea\label{}\eta_1&=&
\frac{2}{1}\left(n_{1}-\frac{1}{2}n_{2}\right)^2+\dots+
\frac{i}{i-1}\left(n_{i-1}-\frac{i-1}{i}n_{i}\right)^2
\nn \\ &&
+\frac{2}{1}\left(n_{s}-\frac{s-1}{2}n_{2}\right)^2+\dots+
\frac{s-i}{s-i-1}\left(n_{i+2}-\frac{s-i-1}{s-i}n_{i+1}\right)^2
\nn \\ &&
+\frac{s-i+1}{s-i}n_{i+1}^2
+\frac{i+1}{i}\left(n_{i}-\frac{i}{i+1}n_{i+1}\right)^2+\frac{s+1}{(i+1)(s-i)}n_{i+1}^2.
\eea
With this formula the integrations over $n_1$, ..., $n_i$ and over $n_{i+1}$, ..., $n_s$ can be done and  renaming $n_i=x$ and $n_{i+1}=y$ we obtain
\bea\label{S2a}S_2&\equiv& \int_{-\infty}^\infty \frac{dn_1}{\sqrt{\pi}}\dots  \int_{-\infty}^\infty \frac{dn_{s} }{\sqrt{\pi}} \
\sum_{i=1}^{s-1} d(n_i,n_{i+1}) \ e^{-\eta_1} \\ 
&=&
\sum_{i=1}^{s-1}\frac{1}{\sqrt{i(s-i)}} \ \int_{-\infty}^\infty \frac{dx }{\sqrt{\pi}} \ \int_{-\infty}^\infty \frac{dy }{\sqrt{\pi}} \
 d(x,y)  
 \nn \\ && ~~~~~~~~~~~
 \cdot \exp\left(-\frac{i+1}{i} \left(x-\frac{i}{i+1}y\right)^2-\frac{s+1}{(i+1)(s-i)} y^2\right).\nn
\eea
The last two integrations and the sum over $i$ give
\be\label{}S_2=-\frac{1}{16 \sqrt{2} (s+1)^2}+\frac{127}{384 \sqrt{2} (s+1)^3}-\frac{103}{192 \sqrt{2} (s+1)^4}+\frac{1}{4 \sqrt{2} (s+1)^5}.
\ee
Taken together the expressions simplify to some extend,
\be S_1+S_2=\frac{1}{48 \sqrt{2} (s+1)^2}-\frac{17}{384 \sqrt{2} (s+1)^3}+\frac{1}{24 \sqrt{2} (s+1)^4} \ .
\ee

Now we calculate the contributions of the double sum in \Ref{Ec6}. Again we separate the first and the last terms,
\bea\label{S3a}S_3&=&
\int_{-\infty}^\infty \frac{dn_1}{\sqrt{\pi}}\dots  \int_{-\infty}^\infty \frac{dn_{s} }{\sqrt{\pi}} 
\sum_{0\le i <  j\le s} b(n_i,n_{i+1},n_j,n_{j+1})
 \nn \\ &=&
\int_{-\infty}^\infty \frac{dn_1}{\sqrt{\pi}}\dots  \int_{-\infty}^\infty \frac{dn_{s} }{\sqrt{\pi}} 
\Big( b(0,n_1,n_s,0)+\sum_{0<j<s} b(0,n_1,n_j,n_{j+1})\nn \\ &&+
\sum_{0<i<s} b(n_i,n_{i+1},n_s,0)
+\sum_{0<i<j<s} b(n_i,n_{i+1},n_j,n_{j+1})   \Big)
\nn \\ &\equiv&  S_{3,1}+S_{3,2}+S_{3,3}+S_{3,4}.
\eea
For $S_{3,1}$ we rewrite $\eta_1$ in the form
\bea\label{}\eta_1&=&
n_1^2+\left(n_2-n_1\right)^2+\left(n_2-n_1-(n_3-n_1)\right)^2+\dots
\nn \\ && ~~~~~~~~~~~~~~~~~~~~~~+
\left(n_{s-1}-n_1-(n_s-n_1)\right)^2+(n_s-n_1)^2\nn \\ &&
-(n_s-n_1)^2+n_s^2
\eea
and apply \Ref{B3} with $u=s-1$, $m_1=n_2-n_1$, ..., $m_u=n_s-n_1$,
\bea\label{}\eta_1&=&
\frac{2}{1}\left(n_2-n_1-\frac{1}{2}(n_3-n_1)\right)^2+\dots+
\frac{s-1}{s-2}\left(n_{s-1}-n_1-\frac{s-2}{s-1}(n_s-n_1)\right)^2\nn \\ &&
+\frac{s}{s-1}\left(n_1-\frac{1}{s}n_s\right)^2+\frac{s+1}{s}n_s^2.
\eea
After integration over $n_2$,..., $n_{s-1}$ and renaming $n_1=x$ and $n_s=y$ we have
\bea\label{}S_{3,1}&=&\frac{1}{\sqrt{s-1}}\int_{-\infty}^\infty \frac{dx}{\sqrt{\pi}}\int_{-\infty}^\infty \frac{dy}{\sqrt{\pi}}
\ b(0,x,y,0) 
\nn \\ && ~~~~~~~~~~~~~~ \cdot
\exp\left(-\frac{s}{s-1}\left(x-\frac{1}{s}y\right)^2-\frac{s+1}{s}y^2\right).
\eea
The last two integrations give
\be\label{S31}S_{3,1}=-\frac{s-1}{8 \sqrt{2} (s+1)^5}.
\ee
For $S_{3,2}$ we rewrite $\eta_1$ in the form
\bea\label{eta31}\eta_1&=&
n_1^2+\left(n_1-n_2\right)^2+\dots+
\left(n_{i-1}-n_i\right)^2+n_i^2\nn \\ &&
-n_i^2+(n_i-n_{i+1})^2-\left(n_{i+1}-n_s\right)^2
\nn \\ &&
+\left(n_{i+1}-n_s\right)^2+\left(n_{i+1}-n_s-(n_{i+2}-n_s)\right)^2
+\dots
\nn \\ && ~~~~~~
+\left(n_{s-2}-n_s-(n_{s-1}-n_s)\right)^2+(n_{s-1}-n_s)^2+n_s^2
\eea
and apply \Ref{B3} to $n_1$, ..., $n_i$ with $u=i$, $m_1=n_1$, ..., $m_u=n_i$ and to $n_{i+1}-n_s$, ..., $n_{s-1}-n_s$ with $u=s-i-1$ and $m_1= n_{i+1}-n_s$, ..., $m_u=n_{s-1}-n_s$ so that $\eta_1$ takes the form
\bea\label{eta31a}\eta_1&=&
\frac{2}{1}\left(n_1-\frac{1}{2}n_2\right)^2+\dots+
\frac{i}{i-1}\left(n_{i-1}-\frac{i-1}{i}n_i\right)^2
 \\ &&
+\frac{2}{1}\left(n_{s-1}-n_s-\frac{1}{2}(n_{s-2}-n_s)\right)^2+\dots
\nn \\ && ~~~~~~~~~~~~~~~~
+\frac{s-i-1}{s-i-2}\left(n_{i+2}-n_s-\frac{s-i-2}{s-i-1}(n_{i+1}-n_s)\right)^2\nn \\ &&
+\frac{i+1}{i}\left(n_i-\frac{i}{i+1}n_{i+1}\right)^2
+\frac{s}{(s-i-1)(i+1)}\left(n_{i+1}-\frac{i+1}{s}n_{s}\right)^2
\nn \\ && 
~~~+\frac{s+1}{s} n_s^2 \ .\nn
\eea
%
After integration over $n_1$,..., $n_{i-1}$ and $n_{i+1}$,..., $n_{s-1}$ and renaming $n_1=x$,  $n_{i+1}=y$ and $n_s=z$ we have
\bea\label{}S_{3,2}&=&\sum_{i=1}^{s-1}\frac{1}{\sqrt{i(s-i-1)}}\int_{-\infty}^\infty \frac{dx}{\sqrt{\pi}}\int_{-\infty}^\infty \frac{dy}{\sqrt{\pi}}\int_{-\infty}^\infty \frac{dz}{\sqrt{\pi}}
\ b(x,y,z,0)  \\ &&
\cdot \exp\left(-\frac{i+1}{i}\left(x-\frac{i}{i+1}y\right)^2
-\frac{s}{(s-i-1)(i+1)}\left(y-\frac{i+1}{s}z\right)^2-\frac{s+1}{s}z^2\right).\nn
\eea
The last three integrations and the sum give
\be\label{S32}S_{3,2}=-\frac{(s-1)^2}{8 \sqrt{2} (s+1)^4} \ .
\ee

In the same way  we calculate $S_{3,3}$. Here we rewrite $\eta_1$ in the form
\bea\label{eta330}\eta_1&=&
n_1^2+\left(n_2-n_1\right)^2+\left(n_{2}-n_1-(n_{3}-n_1)\right)^2\dots
\nn \\ && ~~~~~~~~~~~~~~~~~~~~~~~~~~+
\left(n_{j-1}-n_1-(n_{j}-n_1)\right)^2+\left(n_{j}-n_1\right)^2+\nn \\ &&
-(n_j-n_1)^2+(n_j-n_{j+1})^2-n_{j+1}^2
\nn \\ &&
+n_{j+1}^2+\left(n_{j+1}-n_{j+2})\right)^2
+\dots+n_s^2
\eea
and apply \Ref{B3} to $n_2$, ..., $n_{j-1}$ with $u=j-1$, $m_1=n_2-n_1$, ..., $m_u=n_{j-1}-n_1$ and to $n_{j+1}$, ..., $n_{s}$ with $u=s-j$ and $m_1= n_{s}$, ..., $m_u=n_{j+1}$ so that $\eta_1$ takes the form
\bea\label{eta33}\eta_1&=&
\frac{2}{1}\left(n_{2}-n_1-\frac{1}{2}(n_{3}-n_1)\right)^2+\dots+
+\frac{j-1}{j-2}\left(n_{j-1}-n_1-\frac{j-2}{j-1}(n_{j}-n_1)\right)^2
\nn \\ && 
+\frac{2}{1}\left(n_s-\frac{1}{2}n_{s-1}\right)^2+\dots+
\frac{s-j}{s-j-1}\left(n_{j+2}-\frac{s-j-1}{s-j}n_{j+1}\right)^2
\nn \\ &&
+\frac{s+1}{s}n_1^2
-\frac{s}{(s-j+1)(j-1)}\left(n_{j}-\frac{s-j+1}{s}n_{1}\right)^2
\nn \\ && ~~~~~~~~~-\frac{s-j+1}{s-j}\left(n_{j+1}-\frac{s-j}{s-j+1}n_{j}\right)^2\ .
\eea
After integration over $n_2$,..., $n_{j-1}$ and over $n_{j+2}$,..., $n_{s}$  and renaming $n_1=x$,  $n_{j}=y$ and $n_s=z$ we have
\bea\label{}S_{3,3}&=&\sum_{j=1}^{s-1}\frac{1}{\sqrt{(j-1)(s-j)}}\int_{-\infty}^\infty \frac{dx}{\sqrt{\pi}}\int_{-\infty}^\infty \frac{dy}{\sqrt{\pi}}\int_{-\infty}^\infty \frac{dz}{\sqrt{\pi}}
\ b(0,x,y,z)  \\ &&\nn
\cdot \exp\left(-\frac{s+1}{s} x^2
-\frac{s}{(s-j+1)(j-1)}\left(y-\frac{s-j+1}{s}x\right)^2
\right. \nn \\ && \left.
-\frac{s-j+1}{s-j}\left(z-\frac{s-j}{s-j+1}y\right)^2\right).
\eea
The last three integrations and the sum give
\be\label{S33}S_{3,3}=-\frac{s-1}{8 \sqrt{2} (s+1)^4}\ .
\ee

The last step is $S_{3,4}$, which in fact contains the double sum. Here we represent $\eta_1$ in the form
\bea\label{eta340}\eta_1&=&
n_1^2+\left(n_1-n_2\right)^2+\dots+\left(n_{i-1}-n_i\right)^2+n_i^2
\nn \\ &&
-n_i^2+(n_i-n_{i+1})^2
\nn \\ &&
+(n_{i+2}-n_{i+1})^2+(n_{i+2}-n_{i+1}-(n_{i+3}-n_{i+1}))^2+\dots
\nn \\ && ~~~~~~~~~~~~~~~~~~~+(n_{j-1}-n_{i+1}-(n_j-n_{i+1}))^2+(n_j-n_{i+1})^2
\nn \\ &&
-(n_j-n_{i+1})^2+(n_j-n_{j+1})^2-n_{j+1}
\nn \\ &&
+n_{j+1}^2+\left(n_{j+1}-n_{j+2})\right)^2+\dots+n_s^2.
\eea
We apply \Ref{B3} three times, first, with $u=i$ and $m_1=n_1$, ..., $m_u=n_i$, second, with $u=j-i-1$ and $m_1=n_{i+2}-n_{i+1}$, ..., $m_u=n_j-n_{i+1}$, and, third, with $u=s-j$ and $m_1=n_s$, ..., $m_u=n_{j+1}$. So we write $\eta_1$ in the form
\bea\label{eta34}\eta_1&=&
\frac{2}{1}\left(n_1-\frac{1}{2}n_2\right)^2+\dots+
\frac{i}{i-1}\left(n_{i-1}-\frac{i-1}{i}n_i\right)^2
\nn \\ &&
+\frac{2}{1}\left(n_{i+2}-n_{i+1}-\frac{1}{2}(n_{i+3}-n_{i+1})\right)^2+\dots+
\nn \\ && ~~~~~~~~~
+\frac{j-i-1}{j-i-2}\left(n_{j-1}-n_{i+1}-\frac{j-i-2}{j-i-1}(n_{j}-n_{i+2}\right)^2
\nn \\ &&
+\frac{2}{1}\left(n_{s}-\frac{1}{2}n_{s-1}\right)^2+\dots+
\frac{s-j}{s-j-1}\left(n_{j+2}-\frac{s-j-1}{s-j}n_{j+1}\right)^2
\nn \\ &&
+\frac{i+1}{i}\left(n_i-\frac{i}{i+1}n_{i+1}\right)^2
+\frac{j}{(i+1)(j-i-1)}\left(n_{i+1}-\frac{i+1}{j}n_j\right)^2
\nn \\ &&~~~~~~~~~~~
+\frac{s+1}{j(s-j+1)}n_j^2
+\frac{s-j+1}{s-j}\left(n_{j+1}-\frac{s-j}{s-j+1}n_j\right)^2 \ .
\eea
All integration except for four, whose variables we rename for $n_i=x$, $n_{i+1}=y$, $n_j=z$ and $n_{j+1}=w$, can be done,
\bea\label{S34}S_{3,4}&=&\sum_{0<i<j<s}
\frac{1}{\sqrt{(j-i-1)(s-j)i}}
\nn \\ &&
\int_{-\infty}^\infty \frac{dx}{\sqrt{\pi}}\int_{-\infty}^\infty \frac{dy}{\sqrt{\pi}}\int_{-\infty}^\infty \frac{dz}{\sqrt{\pi}}\int_{-\infty}^\infty \frac{dw}{\sqrt{\pi}}
\ b(x,y,z,w) \nn \\ &&
  \cdot \exp\Bigg[
-\frac{i+1}{i} \left(x-\frac{i}{i+1}y\right)^2
-\frac{j}{(j-i+1)(i+1)}\left(y-\frac{i+1}{j}z\right)^2
\nn \\ &&~~~~~~
-\frac{s+1}{(s-j+1)j}z^2
-\frac{s-j+1}{s-j}\left(w-\frac{s-j}{s-j+1}z\right)^2\Bigg].
\eea
The last integrations   give
\be\label{}S_{3,4}=-\sum_{0<i<j<s}\frac{(2 i+1) (4 j-3 s-1)}{8 \sqrt{2} (s+1)^5}
\ee
and after the summations over $i$ and $j$
\be\label{}S_{3,4}=\frac{(s-2) (s-1) (5 s+3)}{48 \sqrt{2} (s+1)^5}\ .
\ee
Together we get for $S_3$, \Ref{S3a},
\be\label{}S_3=-\frac{(s-1) s}{48 \sqrt{2} (s+1)^4}\ .
\ee
Collecting together all parts we obtain finally
\be\label{C31}S_1+S_2+S_3=\frac{7}{384 \sqrt{2} (s+1)^3}\ ,
\ee
which is the result quoted in Eq.\Ref{longint}.

\bibliographystyle{unsrt}\bibliography{../../../Literatur/Bordag,../../../Literatur/libri,../../../Literatur/articoli,../SpherePlane}

   \end{document}